\documentclass[11pt]{article}
\usepackage{iap2000,epsfig}

\def\be{\begin{equation}}
\def\ee{\end{equation}}
\def\bea{\begin{eqnarray}}
\def\eea{\end{eqnarray}}

\def\etal{{\it et al. }}
\begin{document}
\vspace*{4cm}
\title{NON THERMAL EMISSION FROM GALAXY CLUSTERS: RADIO HALOS}

\author{L. Feretti$^1$, G. Brunetti$^{1,2}$, G. Giovannini$^{1,3}$, 
F. Govoni$^{1,2}$, G. Setti$^{1,2}$ }

\address{$^1$ Istituto di Radioastronomia CNR, Bologna, Italy\\
$^2$ Dipartimento di Astronomia, Univ. Bologna, Italy\\
$^3$ Dipartimento di Fisica, Univ. Bologna, Italy}

\maketitle\abstracts{
 The number of diffuse radio halos in clusters of
galaxies has grown in recent years, making it possible to derive
statistical properties of these sources and of the hosting clusters.
We show that diffuse sources are associated with X-ray luminous 
clusters which have undergone recent merger processes. 
The radio and X-ray structures are often similar, and  correlations
are found between radio and X-ray parameters. A model for the
formation and maintenance of these sources is suggested, including a first
phase of relativistic particle injection 
by past major merger events, starbust and AGN activity
and a second phase of reacceleration 
by the energy supplied  from shocks and turbulence in a recent merger event.
}

\section{Introduction}

It is well known that the radio emission from clusters of galaxies 
generally originates from individual radio emitting galaxies.
In addition,  synchrotron
emission associated with  the intergalactic medium, rather than 
to a particular galaxy, may be detected in some cases.
Clusters of galaxies may exhibit diffuse extended radio sources,
which have been classified as radio halos, relics and mini-halos
(Feretti \& Giovannini 1996). 
The radio halos  are the most spectacular expression
of cluster non-thermal emission.  
They permeate the cluster centers with size of the order of a Mpc 
or more. They  are characterized by low surface brightness,  steep
radio spectrum, and no polarized emission  detected.

The importance of these sources is that they represent large scale 
features, which are related to other cluster
properties in the optical and X-ray domain, and are
thus directly connected to the cluster history
and evolution. 

 In this paper, the observational properties  of radio halos are summarized, 
together with the properties of their parent clusters and
a  model of radio 
halo formation and evolution is described.
Intrinsic parameters are calculated with H$_0$ = 50 km s$^{-1}$ Mpc$^{-1}$
and q$_0$ = 0.5.

\begin{figure}
\psfig{figure=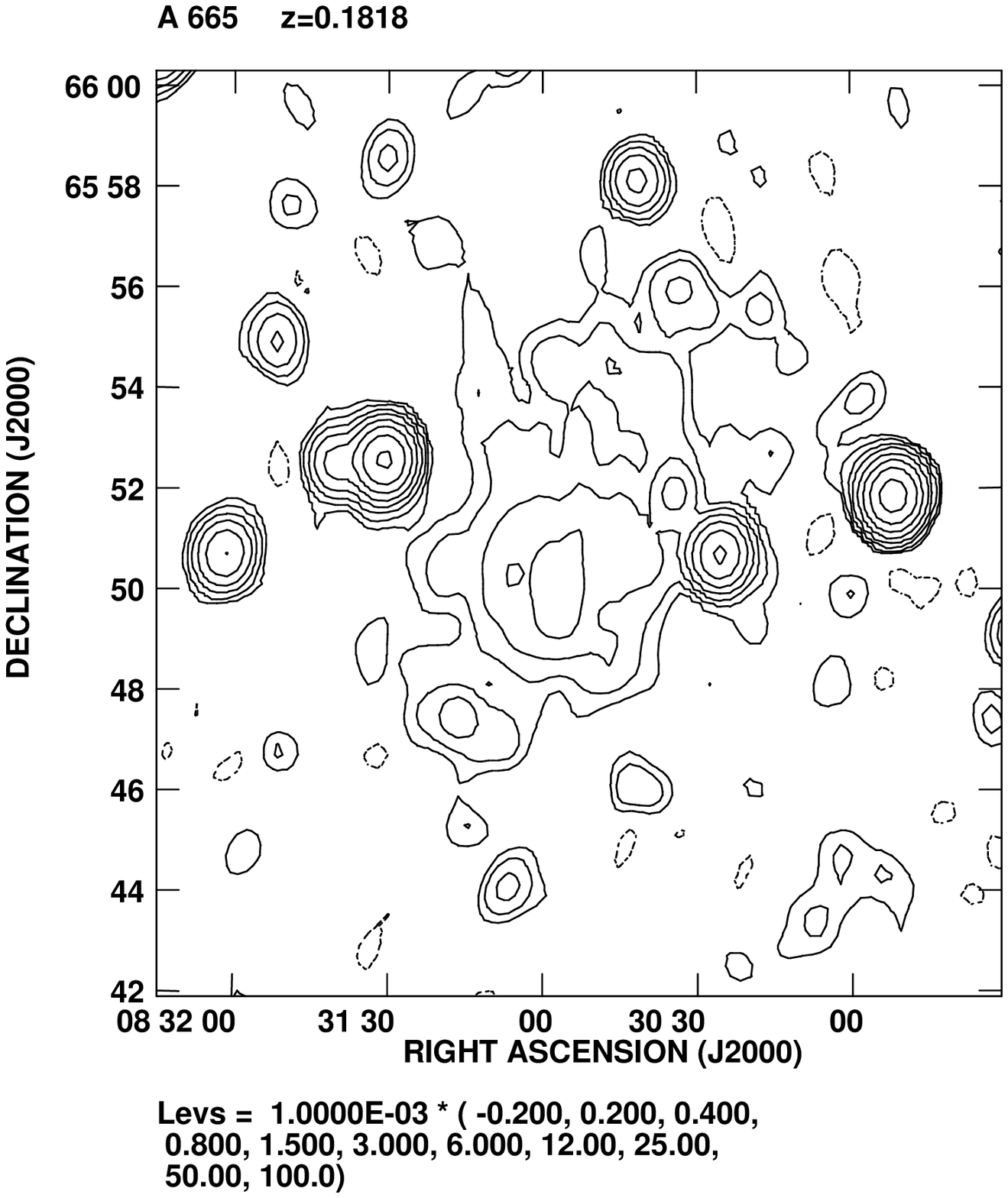,height=3in}
\psfig{figure=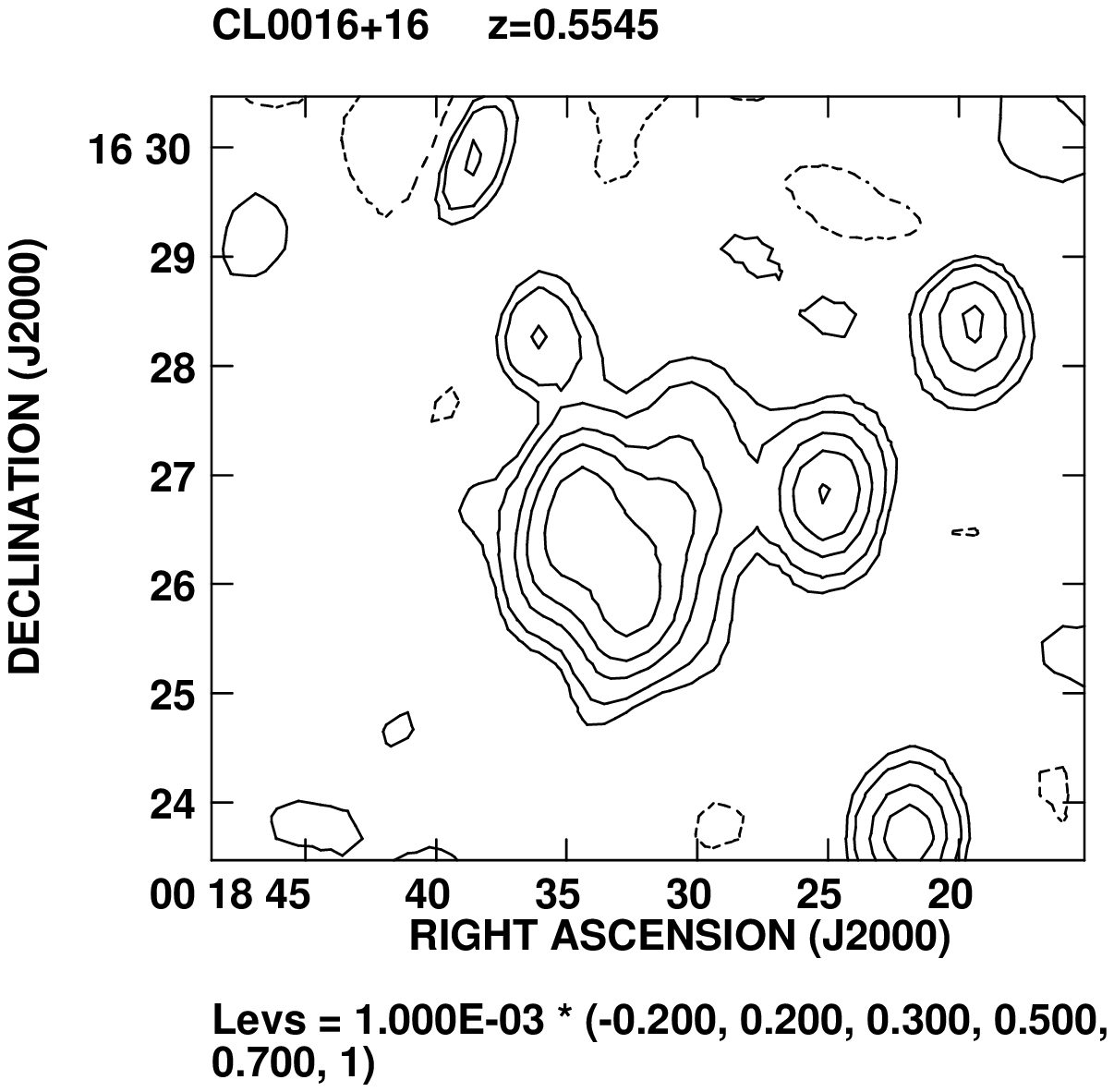,height=3in}
\caption{{\bf Left panel}: 
Image of the giant radio halo  detected at the center of A~665.
The HPBW is 42$^{\prime\prime}$ $\times$ 52$^{\prime\prime}$ (RA$\times$DEC).
{\bf Right panel}: Image of the radio halo present in the central 
region of the distant cluster CL~0016+16. 
The HPBW is 45$^{\prime\prime}$ $\times$ 60$^{\prime\prime}$ (RA$\times$DEC).}
\end{figure}

\section{Examples and properties of radio halos}

The diffuse source Coma C in the Coma cluster, 
discovered  30 years ago (Willson 1970), is the 
prototypical example of a cluster {\it radio halo}.
It is located at  the cluster center, it
has a steep radio spectrum ($\alpha \sim
1.3$) and is extended $\sim$ 1 Mpc (Giovannini \etal 1993). 
Other well studied radio halos are detected in the clusters
A~2255 (Feretti \etal 1997a)  and A~2319 (Feretti \etal 1997b).

In Fig. 1 and Fig. 2, we show images of giant and powerful
radio halos discovered in distant clusters. 
The radio images presented here have been 
obtained with the Very Large Array at 20 cm.
The radio halo in A~665 is  one of the largest known
so far, with the size of 2.4 Mpc (Giovannini \& Feretti 2000). 
The radio emission is slightly asymmetric with respect to the cluster center,
being brighter and more extended toward NW.
This cluster has a temperature of 8.3 keV, no cooling flow, 
and is suggested by X-ray data to be in 
a postmerger state  (Markevitch 1996, Gomez \etal 2000).
The halo in CL~0016+16 (Giovannini \& Feretti 2000) is 
the most distant radio halo known so far. It is extended 1.1 Mpc and
is co-located with the cluster X-ray emission. 
The cluster has a temperature of 
8.2 keV, and shows strong
evidence of substructure which is possibly due to merging of a galaxy
group with the main cluster (Neumann \&  B\"ohringer 1997).
The halo in A~520 is extended 1.4 Mpc, with the radio
structure elongated in NE-SW direction (Govoni {\it et al.}, in preparation).
The two tailed radio sources on the eastern side are cluster radio galaxies.
The cluster temperature is 8.6 keV. 
The halo in A~2744 shows a regular and symmetric structure, 
with a total size of about 2.2 Mpc (Govoni {\it et al.}, in preparation).
The elongated diffuse emission in the  NE peripheral region is classified
as a cluster relic.
This cluster is very hot, with a temperature of 11 keV, and
shows  X-ray substructure, which is possible indication of a recent merger. 

In general, the  sizes of radio halos are  typically larger than 1 Mpc. 
Their radio powers are of the order of 
10$^{24}$-10$^{25}$ W Hz$^{-1}$ at 1.4 GHz.
Minimum energy densities are between
$\sim$ 5 $ $10$^{-14}$ and 2 10$^{-13}$ erg cm$^{-3}$.
This implies that the pressure of relativistic electrons is much lower
than that of the thermal plasma.
Equipartition magnetic fields are about $\sim$0.1-1$\mu$G.
These values are lower than the cluster magnetic field strenghts obtained
from Rotation Measure arguments, and are of the same order as the estimates
 derived from Inverse-Compton X-ray emission. It should be noticed that the
Rotation Measure estimates could be sensitive to the presence of filamentary 
sructure in the cluster and/or to the existence of local turbulence around
the radio galaxies, and could therefore be higher 
than the average cluster value.

\begin{figure}
\psfig{figure=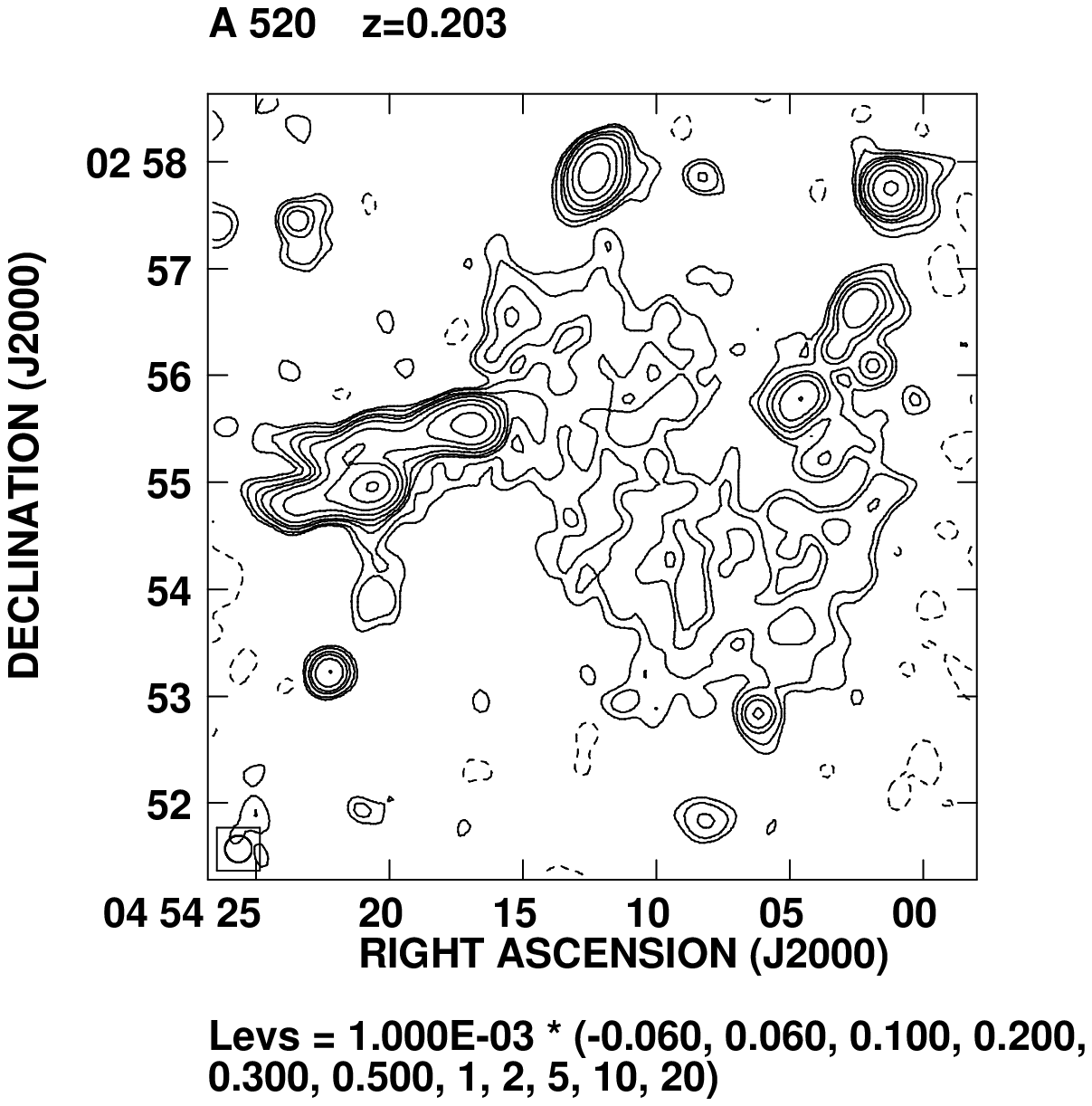,height=3in}
\psfig{figure=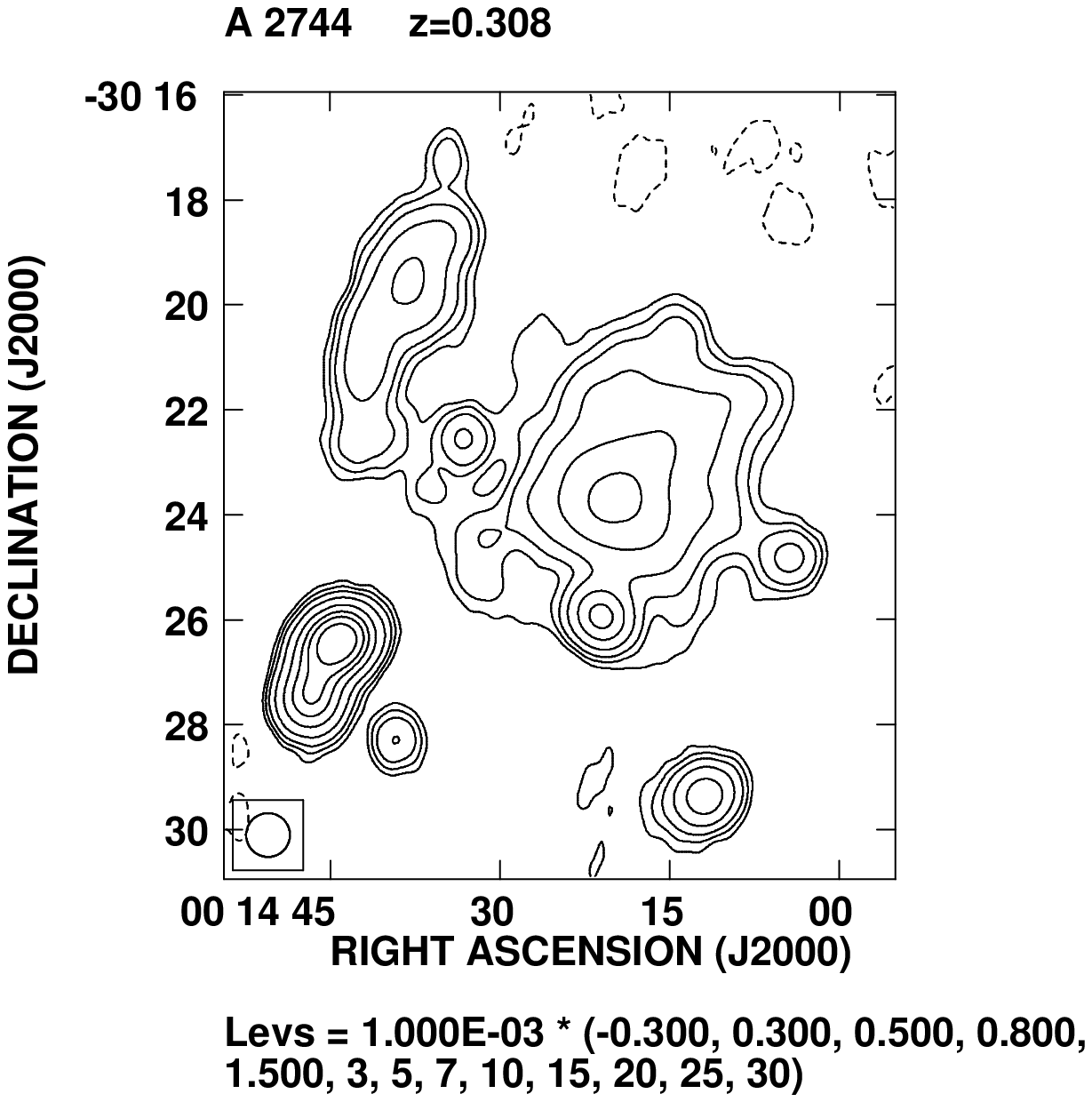,height=3in}
\caption{ {\bf Left panel}: The radio halo in A~520, imaged
with a circular synthesized restoring beam 
with  FWHM = 15$^{\prime\prime}$. {\bf Right panel}: 
Image of the radio halo in  A~2744. The elongated emission  
in the NE peripheral region is a relic.
The synthesized restoring beam is a
circular Gaussian with a FWHM of 50$^{\prime\prime}$. 
}
\end{figure}

\section{Properties of clusters with radio halos}

Unlike the thermal X-ray emission,
the radio halos are not a common feature in clusters of galaxies.
Until recently,  the number of halos  was small, thus
these sources were considered to be a rare phenomenon.
Nowadays, thanks to the better sensitivity of radio telescopes
and to the existence of deep surveys, 
more than 25 clusters are known to host radio halos at their centers
(Giovannini \etal 1999), 
therefore some statistical considerations can be drawn.

The percentage of clusters showing central radio halos in the complete 
X-ray flux limited sample extracted from 
 Ebeling \etal (1996) is 
of $\sim$5\%.  The detection rate
increases with the X-ray luminosity 
being  $\sim$30-35\%  in clusters with X-ray
luminosity  larger than 10$^{45}$ erg s$^{-1}$.
The clusters hosting a radio halo have a
significantly higher X-ray luminosity than
clusters without a diffuse source ($>$ 99.9\%
confidence level with a KS test) (Giovannini \etal 2000).

The high luminosity of  clusters with radio halos implies 
that these clusters also have a high temperature and
a large mass. This is consistent with the serendipitous detection
of halos during the attempts
to detect  Sunyaev-Zeldovich effect in massive high redshift clusters. 

In previous studies (e.g. Feretti 1999, Feretti 2000) radio halos have 
been found to be  associated with clusters showing
indication of merger processes from X-ray and optical structure, 
and from X-ray temperature gradients.

Moreover, a correlation is found between the monochromatic radio power 
at 1.4 GHz of radio halos and the bolometric 
X-ray luminosity of the parent clusters. This 
implies a  correlation between radio power
and cluster temperature,  as shown by Liang \etal (2000).
Since the cluster X-ray luminosity and mass are correlated as well as
the temperature and mass, it follows that the
halo radio power also correlates with the cluster mass. 

Radio structures of halos show in many cases close similarity to 
the X-ray structures, suggesting
a causal connection between the hot and relativistic plasma.
To quantify the similarity in the radio and X-ray
structures, Govoni \etal (2000) performed an analysis of the
radio and the X-ray
emission in four clusters of galaxies containing radio halos 
(Coma, A~2255, A~2319, A~2744).
This study leads to a correlation between the radio and the X-ray 
brightness in all the analyzed clusters: a higher X-ray brightness 
is associated with a higher radio brightness. The relation between 
the radio and the X-ray
brightness is found to be linear in A~2255 and A~2744, whereas
it is represented by a sublinear power-law in the other two 
clusters of galaxies.

\section {The 2-phase model for radio halo formation and evolution}

Radio halos directly demonstrate the existence of
cluster--wide magnetic fields and of relativistic
electrons within the cluster intergalactic medium.
A number of models have been invoked to explain 
radio halo formation and their broad band radiative
properties (see En{\ss}lin 2000; Sarazin  this meeting). 
Observations suggest that central radio halos are 
strictly related to the presence of recent merging 
processes, which can provide the energy for the electron
reacceleration 
and magnetic field amplification.
Furthermore, as discussed above, also the high X--ray luminosity, 
the large cluster mass and/or the high cluster temperature 
appear to be necessary conditions 
for their formation. 
One would conclude that the dynamical history of the clusters 
is crucial to trigger a radio halo. 
We have developed a {\it two phase} model
consisting of a {\it first phase} during which 
relativistic electrons are injected in the cluster volume
by strong shocks, sturburst and/or AGN activity and of
a {\it second phase} during which the 
aged
electrons are reaccelerated
by recent merging processes (Brunetti \etal 2000a).
In the framework of this model we find two general results:

$\bullet$
The formation of luminous radio halos may not be a common phenomenon.
Indeed the injected relativistic electrons suffer efficient
radiation and Coulomb losses and rapidly cool.
This prevents the formation of a radio halo if the time
gap between the {\it first} and {\it second phase} $\Delta t$
is larger than $\sim 2-3$ Gyr.
Furthermore, in order to allow the formation of a radio halo
during the
reacceleration {\it phase}, 
the number of relativistic electrons injected during the {\it first phase} 
should be large enough, increasing with increasing $\Delta t$.
In this framework one could claim that the radio halo 
formation is favoured in the
case of massive clusters which probably derive from a stronger
merger activity in the past and where 
the injection of larger quantities of cosmic rays  
is more efficient.

\begin{figure}
\centerline{\psfig{figure=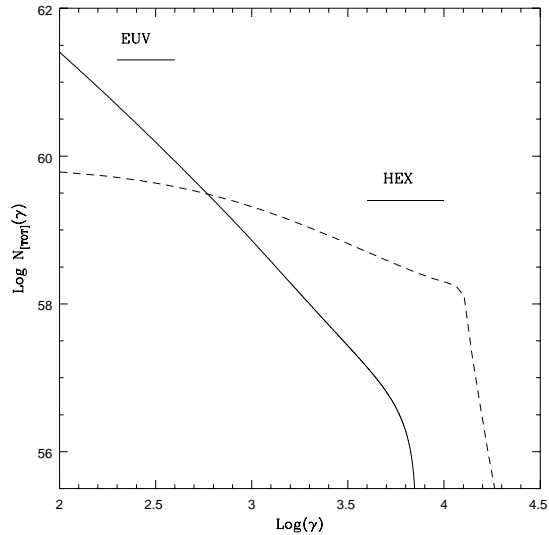,height=3in}}
\caption{
Plot of the energy distribution of the relativistic electron populations
in the Coma cluster. The dashed line represents the main electron population,
integrated over the cluster volume, injected during the 
{\it first phase} and reaccelerated during the 
{\it second phase}. The solid line refers to the 
 additional electron population
injected by the radio galaxy NGC~4869.
The typical energies of the electrons emitting via IC in
the hard X--ray band (HEX) and in the EUV band are also
indicated.
\label{fig:radi}}
\end{figure}

$\bullet$
Central radio halos triggered during the {\it second phase}
by diffuse reacceleration processes are expected 
to show a steepening in the synchrotron spectrum
with increasing distance from the center of the cluster.
Indeed,
the magnetic field strength is expected to be
a decreasing function of the radius and so probably also
the reacceleration efficiency. 

The model has been applied in detail to the well studied radio halo
Coma C (Brunetti \etal 2000a).
The radial  steepening of the radio spectrum observed in Coma C
(Giovannini \etal 1993, Deiss \etal 1997) 
has been used to constrain the
physical conditions in the cluster, obtaining 
reacceleration efficiencies of the order of $10^{-8}$yr$^{-1}$
and average magnetic field strengths ranging from 1--3 $\mu G$ 
in the central regions to 0.05--0.1 $\mu G$ in the external parts 
 of the cluster ($\sim$ 2--3 Mpc).
The model satisfactorily reproduces the total radio spectrum of Coma C and 
the size of the halo. The expected
hard X--ray inverse Compton emission, mainly     
produced at relatively large distances from
the cluster center ($\geq 1.5$ Mpc), is consistent with the  
flux detected by BeppoSAX (Fusco--Femiano \etal 1999).

As previously discussed, the relativistic electrons radiating 
within the halo were injected during the  past history of the cluster. 
However, additional electron injection could also derive 
from  recent activity in the cluster.
Although it has  been shown that the radio
halo phenomenon does not appear to be correlated with the presence 
of AGNs in the cluster (in particular with head tail radio sources,
Giovannini \& Feretti 2000), even a modest injection of 
star forming galaxies and AGNs could affect 
the broad band properties of radio halos.
In a recent paper, Brunetti \etal (2000b) have calculated, in the framework
of the above described {\it two phase} model,  
the evolution of the spectrum of relativistic electrons 
recently injected (in the last 0.1--0.3 Gyr)
by the head tail radio galaxy NGC 4869 in Coma. 
It was found that the fresh--injected population does not significantly
contribute to the $\geq 300$ MHz spectrum of the radio halo 
but it could significantly contribute to the emission at lower frequencies.
Accurate measurements at low frequencies, such as those obtainable with the
VLA at 74 MHz, are of crucial importance to test the model.
In this scenario, a large fraction (if not all) of the 
EUV excess detected in the Coma cluster (Bowyer \etal 1999)
may be accounted for by the inverse Compton scattering of CMB
photons by the recently injected population, whose  contribution
to the hard X--ray emission is however not relevant.
The spectral distribution of the two electron populations   
in the Coma cluster are reported in Fig.3.
Due to radiative losses and reacceleration gains, the energy
distribution  of the recently injected population is expected to 
become very similar to the main electron population 
with increasing time ( $\geq 1$ Gyr).

\section*{Acknowledgments}

This work was partly supported by the Italian Space Agency (ASI).

\end{document}